\documentstyle[epsfig,aps,prd,twocolumn]{revtex}

\def\ave#1{{\left\langle #1 \right\rangle}}
\def\etal{{\it et al.}}

\def\ie{{\it i.e.}}

 \def\lsim{\mathrel {\vcenter {\baselineskip 0pt \kern 0pt
    \hbox{$<$} \kern 0pt \hbox{$\sim$} }}}
    \def\gsim{\mathrel {\vcenter {\baselineskip 0pt \kern 0pt
    \hbox{$>$} \kern 0pt \hbox{$\sim$} }}}
\def\vbar{{\overline v}}

\def\gammabar{{\overline \gamma}}

\begin{document}

\title{Classical Radiation Processes 
in the Weizs\"acker-Williams Approximation}

\author{Max S.~Zolotorev}
\address{Center for Beam Physics, Lawrence Berkeley National Laboratory,
Berkeley, CA 94720}

\author{Kirk T.~McDonald}
\address{Joseph Henry Laboratories,
Princeton University, Princeton, NJ 08544}

\date{August 25, 1999}

\maketitle

\begin{abstract}
The main features of radiation by relativistic electrons 
are well approximated in 
the Weizs\"acker-Williams method of virtual quanta. 
This method is best known for its application to
radiation during elementary particle collisions, but is equally useful 
in describing ``classical'' radiation emitted during the interaction of a 
single relativistic electron
with an extended system, such as synchrotron radiation, undulator radiation,
transition radiation and \v Cerenkov radiation.

\end{abstract}

\vspace{0.25in}

\section{The Weizs\"acker-Williams Approximation}

Following an earlier discussion by Fermi \cite{Fermi},
Weizs\"acker \cite{Weizsacker} and Williams \cite{Williams}
 noted that the electromagnetic fields of an 
electron in uniform relativistic motion are predominantly transverse, 
with ${\bf E} \approx {\bf B}$
(in Gaussian units).  This is very much like the fields of a plane wave, so one
is led to regard a fast electron as carrying with it a cloud of virtual
photons that it can shed (radiate) if perturbed.  

The key feature of the frequency spectrum of the fields can be estimated as
follows.  To an observer at rest at
 distance $b$ from the electron's trajectory, the peak electric
field is $E = \gamma e/b^2$, and the field remains above half this strength
for time $b/\gamma c$, so the frequency 
spectrum of this pulse extends up to $\omega_{\rm max} \approx \gamma c/b$.  
The total energy of the pulse (relevant to this observer) is 
$U \approx E^2 {\rm Vol} \approx \gamma^2 e^2/b^4 \cdot b^2 \cdot b/\gamma
\approx \gamma e^2/b$.

If the electron radiates all of this energy, the energy spectrum would be
\begin{equation}
{dU(\omega) \over d\omega} \approx {U \over \omega_{\rm max}}
\approx {e^2 \over c}.
\label{eq62}
\end{equation}
This result does not depend on the choice of impact parameter  $b$, and is 
indeed of general validity (to within a factor of $\ln \gamma$).  The number of
photons $n_\omega$ of frequency $\omega$ is thus
\begin{equation}
dn_\omega = {dU(\omega) \over \hbar\omega} 
\approx {e^2 \over \hbar c} {d\omega \over \omega}
= \alpha {d\omega \over \omega},
\label{eq63}
\end{equation}
where $\alpha = e^2/\hbar c \approx 1/137$ is the fine structure constant.

The quick approximation (\ref{eq62})-(\ref{eq63}) is not accurate at high 
frequencies.
In general, additional physical arguments are needed to identify the
maximum frequency of its validity, called the characteristic or critical
frequency $\omega_C$,  or equivalently, the minimum relevant
impact parameter $b_{\rm min}$. 
A more detailed evaluation of the high-frequency tail of the
virtual photon spectrum shows it to be \cite{Fermi,Weizsacker,Williams,Jackson}
\begin{equation}
dn_\omega \approx  \alpha {d\omega \over \omega} 
e^{-2\omega b_{\rm min}/\gamma c} \qquad \mbox{(high frequency)}.
\label{eq63a}
\end{equation}
From this, we see the general relation between the critical frequency and
the minimum impact parameter is
\begin{equation}
\omega_C \approx \gamma {c \over b_{\rm min}}, \qquad
b_{\rm min} \approx \gamma \lambda_C.
\label{eq63b}
\end{equation}

The characteristic angular spread $\theta_C$  of the radiation pattern near the
critical frequency can be estimated from eq.~(\ref{eq63b}) by noting that
the radiation is much like that of a beam of light with waist $b_{\rm min}$.
Then, from the  laws of diffraction we conclude that
\begin{equation}
\theta_C \approx {\lambda_C \over b_{\rm min}} \approx {1 \over \gamma}.
\label {eq63d}
\end{equation}
This behavior is also expected in that a ray of light emitted in the
electron's rest frame at $90^\circ$ appears at angle $1/\gamma$  to the 
laboratory direction of the electron. 

\subsection{The Formation Length}

To complete an application of the Weizs\"acker-Williams method, 
we must also know over what
interval the virtual photon cloud is shaken off the electron to become
the radiation detected in the laboratory.
Intense (and hence, physically interesting)
radiation processes are those in which the entire cloud of virtual photons
is emitted as rapidly as possible.
This is usefully described by 
the so-called formation time $t_0$ and the corresponding formation length
$L_0 = vt_0$ where $v \approx c$ is the velocity of the relativistic electron.

The formation length (time)
is the distance (time) the electron travels while a radiated wave advances one
wavelength $\lambda$ ahead of the projection of the electron's motion onto
the direction of observation.  The wave takes on the character of
radiation that is no longer tied to its source only after the formation time
has elapsed.
That is, 
\begin{equation}
\lambda = c t_0 - v t_0 \cos\theta
\approx L_0 (1 - \beta \cos\theta) 
\approx L_0 \left( {1 \over 2 \gamma^2} + {\theta^2 \over 2} \right),
\label{eq64}
\end{equation}
for radiation observed at angle $\theta$ to the electron's trajectory.
Thus, the formation length is given by
\begin{equation}
L_0 \approx {2 \lambda \over \theta^2 + 1/\gamma^2}
\label{eq65}
\end{equation}
If the frequency of the radiation is near the critical frequency (\ref{eq63b}),
then the radiated intensity is significant only for $\theta \lsim \theta_C
\approx 1/\gamma$, and the formation length is
\begin{equation}
L_0 \approx \gamma^2 \lambda   \qquad \qquad (\lambda \approx \lambda_C).
\label{eq65a}
\end{equation}

A good discussion of the formation length in both classical and quantum
contexts has been given in ref.~\cite{Klein}.

\subsection{Summary of the Method}

A relativistic electron carries with it a virtual photon spectrum of
$\alpha$ photons per unit frequency interval.  When radiation occurs, for
whatever reason, the observed frequency spectrum will closely follow this
virtual spectrum.  In cases where the driving force for the radiation
extends over many formation lengths, the spectrum of radiated photons per 
unit path
length for intense processes is given by expressions (\ref{eq63})-(\ref{eq63a}),
which describe the radiation emitted over one formation length,
divided by the formation length (\ref{eq65}):
\begin{equation}
{dn_\omega \over dl} \approx {\alpha \over L_0(\omega)} {d\omega \over \omega}
\times \left\{
\begin{array}{llll}
1 & & & (\omega < \omega_C), \\
e^{-\omega/\omega_C} & & & 
(\omega \geq \omega_C).
\end{array} \right.
\label{eq63c}
\end{equation}

Synchrotron radiation, undulator radiation, transition radiation, and
\v Cerenkov radiation are examples of processes which can be described
within the context of classical electromagnetism, but for which 
the Weizs\"acker-Williams approximation is also suitable.  \v Cerenkov
radiation and transition radiation are often thought of as rather weak
processes, but the Weizs\"acker-Williams viewpoint indicates that they are 
actually as intense as is possible for radiation by a single charge, in the
sense that the entire virtual photon cloud is liberated over a formation length.

In this paper, we emphasize a simplified version of the Weizs\"acker-Williams 
method with the goal of illustrating the principle qualitative features of
various radiation processes.  A more detailed analysis can reproduce the
complete forms of the classical radiation, as has been demonstrated for
synchrotron radiation by Lieu and Axford \cite{Lieu}.

\section{Synchrotron Radiation}

Synchrotron radiation arises when a charge, usually an electron,
 is deflected by a magnetic field.
For a large enough region of uniform magnetic field, the electron's
trajectory would be a complete circle.  However, synchrotron radiation as 
described
below occurs whenever the magnetic field region is longer than a formation
length.  The radiation observed when the magnetic field extends for less
than a formation length has been discussed in 
refs.~\cite{Lieu,Coisson,Bagrov}.

\subsection{The Critical Frequency}

An important fact about synchrotron radiation is that
the frequency spectrum peaks near 
the critical frequency, $\omega_C$, which depends on the radius $R$ of
curvature of the electron's trajectory, and on the Lorentz factor $\gamma$ via
\begin{equation}
\omega_C \approx \gamma^3 {c \over R}.
\label{eq60}
\end{equation}
Since $\omega_0 = c/R$ is the angular velocity for
particles with velocity near the speed of light, synchrotron radiation 
occurs at very high harmonics of this fundamental frequency.  The wavelength
at the critical frequency is then
\begin{equation}
\lambda_C \approx {R \over \gamma^3}.
\label{eq61}
\end{equation}

For completeness, we sketch a well-known argument leading to eq.~(\ref{eq60}).
The characteristic frequency $\omega_C$ is the reciprocal of
the pulselength of the radiation from a single electron according to an
observer at rest in the lab.  In the
case of motion in a circle, the electron emits a cone of radiation of
angular width $\theta = 1/\gamma$ 
according to eq.~(\ref{eq63d}) that rotates with angular velocity 
$\omega = c/R$.  Light within this cone reaches the fixed observer during time
interval $\delta t' = \theta/\omega \approx R/\gamma c$.  However, this time 
interval measures the retarded time $t'$ at the source, not the time $t$ at the
observer.  Both $t$ and $t'$ are measured in the lab frame, and are
related by $t' = t - r/c$ where $r$ is the distance between the source and
observer.  When the source is heading towards the observer, we have
$\delta r = -v \delta t'$, so $\delta t = \delta t'(1 - v/c) \approx
\delta t'/2\gamma^2 \approx R/\gamma^3c$, from which eq.~(\ref{eq60}) follows.

\subsection{The Formation Length}

The formation length $L_0$ introduced in eq.~(\ref{eq65}) applies for 
radiation processes during
which the electron moves along a straight line, such as \v Cerenkov radiation
and transition radiation.  But, synchrotron radiation occurs when the 
electron moves in the arc of a circle of radius $R$.  During the formation
time, the electron moves by formation angle $\theta_0 = L_0 / R$ with respect 
to the center of the circle. 
We now reconsider the derivation of the formation time, noting that while the
electron moves on the arc $R\theta_0 = vt_0$ of the circle, the radiation moves 
on the chord $2R\sin(\theta_0/2) \approx R\theta_0 - R\theta_0^3/24$.  Hence,
\begin{eqnarray}
\lambda & = & c t_0 - \mbox{chord} 
\approx {c R \theta_0 \over v} - R\theta_0 + {R\theta_0^3 \over 24}
\nonumber \\
& \approx & R \theta_0(1 - \beta) + {R\theta_0^3 \over 24}
\approx {R \theta_0 \over 2\gamma^2}  + {R\theta_0^3 \over 24},
\label{eq66}
\end{eqnarray}
for radiation observed at small angles to the chord.

For wavelengths longer than $\lambda_C$, the formation angle grows large
compared to the characteristic angle $\theta_C \approx 1/\gamma$,
and the first term of eq.~(\ref{eq66}) can be neglected compared to the second.
In this case,
\begin{equation}
\theta_0 \approx \left( {\lambda \over R} \right)^{1/3}
\approx {1 \over \gamma} \left( {\lambda \over \lambda_C} \right)^{1/3}
\qquad (\lambda \gg \lambda_C),
\label{eq67}
\end{equation}
and \begin{equation}
L_0 \approx R^{2/3} \lambda^{1/3} 
\approx \gamma^2 \lambda_C \left( {\lambda \over \lambda_C} \right)^{1/3}
\qquad (\lambda \gg \lambda_C),
\label{eq67a}
\end{equation}
using eq.~(\ref{eq61}). 

The formation angle $\theta_0(\lambda)$ can
also be interpreted as the characteristic angular width of the
radiation pattern at this wavelength.
A result not deducible from the simplified arguments given above is that
for $\lambda \gg \lambda_C$,
the angular distribution of synchrotron radiation falls off exponentially:
$dU(\lambda)/d\Omega \propto e^{-\theta^2/2\theta_0^2}$.  See, for example,
sec.~14.6 of \cite{Jackson}.  

For wavelengths much less than $\lambda_C$,
the formation length is short, the formation angle is small, 
and the last term of eq.~(\ref{eq66}) can be neglected.  Then, we find that
\begin{equation}
\theta_0 \approx {\lambda \over \gamma \lambda_C},
\qquad
L_0 \approx \gamma^2 \lambda 
\qquad (\lambda \ll \lambda_C),
\label{eq66a}
\end{equation}
the same as for motion along a straight line, eq.~(\ref{eq65a}).
In this limit, our approximation neglects the curvature of the particle's
trajectory, which is an essential aspect of synchrotron radiation, and
we cannot expect our analysis to be very accurate.  But for $\lambda \ll
\lambda_C$, the rate of radiation is negligible.  

Of greater physical 
interest is the region $\lambda \approx \lambda_C$ where the frequency
spectrum begins to be exponentially damped but the rate is still reasonably
high.  The cubic equation (\ref{eq66}) does not yield a simple analytic
result in the region.  So, we interpolate between the limiting results for
$\theta_0$ at large and small wavelengths, eqs.~(\ref{eq67}) and (\ref{eq66a}),
and estimate that
\begin{equation}
\theta_0 \approx {1 \over \gamma} \sqrt{\lambda \over \lambda_C}
\qquad (\lambda \approx \lambda_C),
\label{eq66b}
\end{equation}
which agrees with a more detailed analysis \cite{Jackson}.   The
corresponding formation length $R \theta_0$ is then
\begin{equation}
L_0 \approx \gamma^2 \sqrt{\lambda \lambda_C}
\qquad (\lambda \approx \lambda_C).
\label{eq67b}
\end{equation}


\subsection{Transverse Coherence Length}

The longitudinal origin of radiation is uncertain to within one formation
length $L_0$.  Over this length, the trajectory of the electron is curved, 
so there is an uncertainty in the transverse origin of the radiation as well.
A measure of the transverse uncertainty is the sagitta $L_0^2/8R$, which we 
label $w_0$ anticipating a useful analogy with the common notation for the 
waist of a focused laser beam.  For $\lambda \gg \lambda_C$, we have from
eq.~(\ref{eq67a}),
\begin{equation}
w_0 \approx {L_0^2 \over R}
\approx R^{1/3} \lambda^{2/3} 
\approx \gamma \lambda_C \left( {\lambda \over \lambda_C} \right)^{2/3}
\qquad (\lambda \gg \lambda_C).
\label{eq68}
\end{equation}
The sagitta (\ref{eq68}) is larger than the minimum transverse
length (\ref{eq63b}), so we expect that the full virtual photon cloud is shaken 
off over one formation length.

For $\lambda \gg \lambda_C$, the characteristic angular spread 
(\ref{eq67}) of the radiation obeys
\begin{equation}
\theta_0 \approx {\lambda \over w_0},
\label{eq69}
\end{equation}
consistent with the laws of diffraction.  Hence, the distance $w_0$ of
eq.~(\ref{eq68}) can also be called the transverse coherence length 
\cite{Mandel} of the source of synchrotron radiation.

The analogy with laser notation is also consistent with identifying the
formation length $L_0$ with the Rayleigh range $z_0 = w_0/ \theta_0$, since we 
see that
\begin{equation}
L_0 \approx {\lambda \over \theta_0^2} \approx {w_0 \over \theta_0}.
\label{eq70}
\end{equation}
A subtle difference between the radiation of a relativistic charge and a 
focused laser beam is that the laser beam has a Guoy phase shift 
\cite{Guoy,Siegman} between its waist and the
far field, while radiation from a charge does not.  

For $\lambda \approx \lambda_C$, the sagitta is $L_0^2/R \approx \gamma^2
\lambda$, using eq.~(\ref{eq67b}).  When $\lambda < \lambda_C$, the
characteristic angle $\theta_0$ given by eq.(\ref{eq66b}) is less than
$\lambda$/sagitta, and the sagitta is no longer a good measure of the
transverse coherence length, which is better defined as $\lambda/\theta_0
\approx \gamma \sqrt{\lambda \lambda_C}$.

\subsection{Frequency Spectrum}

The number of
photons radiated per unit path length $l$ during synchrotron radiation is
obtained from the Weizs\"acker-Williams spectrum (\ref{eq63c}) using
eqs.~(\ref{eq67a}) and (\ref{eq67b}) for the formation length:
\begin{equation}
{dn_\omega \over dl}  
\approx \left\{
\begin{array}{llll}
\alpha \omega_C^{2/3} d\omega / \gamma^2 c \omega^{2/3} 
& & & (\lambda \gg \lambda_C), \\
\alpha \omega_C^{1/2} e^{-\omega/\omega_C} d\omega / \gamma^2 c \omega^{1/2} 
& & & (\lambda \lsim \lambda_C).
\end{array} \right.
\label{eq71}
\end{equation}

We multiply by $\hbar\omega$ to recover
the energy spectrum:
\begin{equation}
{dU(\omega) \over dl} 
\approx \left\{
\begin{array}{llll}
e^2  \omega_C^{2/3} \omega^{1/3} d\omega / \gamma^2 c^2 
& & & (\lambda \gg \lambda_C), \\
e^2 \omega_C^{1/2} \omega^{1/2} e^{-\omega/\omega_C} d\omega / \gamma^2 c^2 
& & & (\lambda \lsim \lambda_C).
\end{array} \right.
\label{eq72}
\end{equation}
Thus, the Weizs\"acker-Williams method shows that
the energy spectrum varies as $\omega^{1/3}$ at low frequencies, and as
$\sqrt{\omega} e^{-\omega/\omega_C}$ at frequencies above the critical
frequency $\omega_C = \gamma^3 c/R$.

The total radiated power is estimated from eq.~(\ref{eq72}) using $\omega 
\approx 
d\omega \approx \omega_C \approx \gamma^3 c/R$, and multiplying by $v \approx c$
to convert $dl$ to $dt$:
\begin{equation}
{dU \over dt} \approx {e^2 \gamma^4 c \over R^2}.
\label{eq72a}
\end{equation}
This well-known result is also obtained from the Larmor formula, $dU/dt = 2 e^2 
a^{\star 2} / 3 c^2$, where the rest-frame acceleration is given by
$a^\star = \gamma^2 a 
\approx \gamma^2 c^2/R$ in terms of lab quantities.

\section{Undulator Radiation}

An undulator is a device that creates a region of transverse magnetic field
that whose magnitude oscillates with spatial period $\lambda_0$.  This field
is constant in time, and is usually lies in a transverse plane (although
helical undulators have been built, and are actually somewhat easily to
analyze).  As an electron with velocity $v$
traverses the undulator, its trajectory involves transverse
oscillations with laboratory wavelength $\lambda_0$, and laboratory
frequency $\omega_0 = c/\lambdabar_0$.  The oscillating electron then emits
undulator radiation.

This radiation is usefully described by first transforming to the average
rest frame of the electron, which is done by a Lorentz boost of
$\gamma = 1/\sqrt{ 1 - (v/c)^2}$ in the first approximation.  The undulator
wavelength in this frame is $\lambda^\star = \lambda_0/\gamma$, and the
frequency of the oscillator is $\omega^\star = \gamma \omega_0$.
The electron emits dipole radiation at this frequency in its average rest
frame.  The laboratory radiation is the transform of this radiation.

Thus, undulator radiation is readily discussed as the Lorentz transform of a 
Hertzian dipole oscillator, and the Weizs\"acker-Williams approximation does
not offer much practical advantage here.  However, an analysis of undulator 
radiation can validate the Weizs\"acker-Williams approximation, while also
exploring the distinction between undulator radiation and wiggler radiation.

\subsection{A First Estimate}

The characteristic angle of undulator radiation in the laboratory is $\theta_C
\approx 1/\gamma$, this being the transform of a ray at $\theta^\star =
90^\circ$ to the electron's lab velocity.  The radiation is nearly
monochromatic, with frequency 
\begin{equation}
\omega_C \approx 2 \gamma \omega^\star
= 2\gamma^2 \omega_0,
\label{eq300}
\end{equation}
and wavelength
\begin{equation}
\lambda_C \approx {\lambda_0 \over 2 \gamma^2}.
\label{eq301}
\end{equation}

The formation length, defined as the distance over which radiation
pulls one wavelength ahead of the electron, 
is $L_0 \approx \gamma^2 \lambda \approx \lambda_0$, the undulator period.  
But when the electron advances
one period, it continues to oscillate, and the amplitude of the radiation
emitted during the second period is in phase with that of the first.
Assuming that the radiation from successive period overlaps in space, 
there will be constructive interference which
continues over the entire length of the
undulator.  In this case, the radiation is not clearly distinct from the near 
zone of
the electron until it leaves the undulator.  Hence, the formation length
of undulator radiation is better defined as
\begin{equation}
L_0 = N_0 \lambda_0,
\label{eq302}
\end{equation}
where $N_0$ is the number of periods in the undulator.

The frequency spread of undulator radiation narrows as the number of
undulator periods increases, and
\begin{equation}
{\Delta \omega \over \omega_C} \approx {1 \over N_0}
\label{eq303}
\end{equation}

We now try to deduce the radiated photon spectrum from 
the Weizs\"acker-Williams approximation (\ref{eq63c}).
The constructive interference over the $N_0$ undulator periods implies that the
radiated energy will be $N_0^2$ times that if there were only one period.  So we
multiply eq.~(\ref{eq63c}) by $N_0^2$ to obtain
\begin{equation}
{dn_\omega \over dl} \approx {N_0^2 \alpha \over L_0} {d \omega \over \omega}
\approx {\alpha \over \lambda_0},
\label{eq304}
\end{equation}
in the narrow band (\ref{eq303}) around the characteristic frequency 
(\ref{eq300}).
The radiated power is $v \hbar  \omega_C \approx c \hbar \omega_C$ times 
eq.~(\ref{eq304}):
\begin{equation}
{dU \over dt} \approx {e^2 c \gamma^2 \over \lambda_0^2},
\label{eq304a}
\end{equation}
using eq.~(\ref{eq300}).

This estimate  proves to be reasonable only for that part of the range of
undulator parameters.  To clarify this, we need to examine the electron's
trajectory through the undulator in greater detail.

\subsection{Details of the Electron's Trajectory}

A magnetic field changes the direction of the electron's velocity, but not
its magnitude.  As a result of the transverse oscillation in the undulator, 
the electron's average forward velocity $\vbar$
will be less than $v$.  The boost to the average rest frame is described by
$\gammabar$ rather than $\gamma$.

In the average rest frame, the electron is not at rest, but oscillates
in the electric and magnetic fields $\tilde E \approx \tilde B = 
\gammabar B_0$, where we use the symbol $\tilde{\phantom{a}}$ to indicate 
quantities in the average rest frame.
The case of a helical undulator is actually simpler than that of a linear
one.  For a helical undulator, the average-rest-frame fields are essentially 
those
of circularly polarized light of frequency $\tilde\omega = \gammabar \omega_0$.
The electron
moves in a circle of radius $R$ at this frequency, in phase with the
electric field $\tilde E$, and with velocity $\tilde v$ and associated Lorentz 
factor $\tilde\gamma$, all related by
\begin{equation}
{\tilde\gamma m \tilde v^2 \over R} = \tilde\gamma m \tilde v \tilde\omega 
= e \tilde E.
\label{eq305}
\end{equation}
From this we learn that
\begin{equation}
\tilde\gamma \tilde\beta = {e \tilde E \over m \tilde\omega c} 
\approx {e B_0 \over m \omega_0 c}
\equiv \eta,
\label{eq306}
\end{equation} 
and hence,
\begin{equation}
\tilde\gamma = \sqrt{1 + \eta^2}, \qquad 
\tilde\beta = {\eta \over \sqrt{1 + \eta^2}},
\label{eq307}
\end{equation}
and
\begin{equation}
R =  {\tilde\beta c \over \tilde\omega} 
= {\eta \tilde\lambdabar \over \sqrt{1 + \eta^2}}
= {\eta \lambdabar_0 \over \gammabar \sqrt{1 + \eta^2}}
\label{eq308}
\end{equation}

Thus, the dimensionless parameter $\eta$ describes many features of the
transverse motion of an electron in an oscillatory field.  It is 
a Lorentz invariant, being proportional to the magnitude of the 4-vector
potential.  

For a linear undulator, $\eta$ is usefully defined as
\begin{equation}
\eta = {e B_{0,{\rm rms}} \over m \omega_0 c},
\label{eq309}
\end{equation}
where the root-mean-square (rms) average is taken over one period.  With the 
definition (\ref{eq309}), the rms values of $\tilde\beta$, $\tilde\gamma$ and
$R$ for a linear undulator of strength $\eta$ are also given by 
eqs.~(\ref{eq307})-(\ref{eq308}).

We can now display a relation for $\gammabar$ by noting that in the
average rest frame, the electron's (average) energy is $\tilde\gamma m c^2 =
m \sqrt{1 + \eta^2} c^2$, while its average momentum is zero there.  Hence, on
transforming back to the lab frame, we have
$\gamma m c^2 = \gammabar \tilde\gamma m c^2$, and so
\begin{equation}
\gammabar = {\gamma \over \sqrt{1 + \eta^2}}.
\label{eq310}
\end{equation}

The transverse amplitude of the motion is
obtained from eqs.~(\ref{eq308}) and (\ref{eq310}):
\begin{equation}
R = {\eta \lambdabar_0 \over \gamma} = 2 \eta \gamma \lambdabar_C,
\label{eq311a}
\end{equation}
recalling eq.~(\ref{eq301}).

\subsection{$\eta > 1$: Wiggler Radiation}

The pitch angle of the helical trajectory is
\begin{equation}
\theta \approx \tan\theta = {R \over \lambdabar_0} = {\eta \over \gamma}.
\label{eq311b}
\end{equation}
Since the characteristic angle of the radiation is $\theta_C \approx 1/\gamma$,
we see that the radiation from one period of the oscillation does not
overlap the radiation from the next period unless 
\begin{equation}
\eta \lsim 1.
\label{eq311c}
\end{equation}
Hence, there is no constructive interference and consequent sharpening of  the
frequency spectrum unless condition (\ref{eq311c}) is satisfied.  

For $\eta > 1$, the radiation is essentially the sum of synchrotron radiation
from $N_0$ separate magnets each $\lambda_0$ long, and this case is called
wiggler radiation.

The laboratory frequency of the radiation is now
\begin{equation}
\omega_C \approx 2 \gammabar^2 \omega_0,
\label{eq311}
\end{equation}
rather than eq.~(\ref{eq300}).  However, in the regime of undulator radiation,
(\ref{eq311c}), there is little difference between the two expressions.

Another aspect of wiggler radiation is that for $\eta \gsim 1$ the motion of the
electron in its average rest frame is relativistic, as can be see from
eq.~(\ref{eq307}).  In this case, higher multipole radiation becomes important,
which appears at integer multiples of frequency $\omega^\star$ in the average 
rest frame, and at the corresponding Lorentz transformed frequencies in the 
lab frame.  The total radiated power is still given by eq.~(\ref{eq304a}), 
so the amount of power
radiated any particular frequency is less than when $\eta \approx 1$.

\subsection{$\eta < 1$: Weak Undulators}

The estimate (\ref{eq304a}) for the power of undulator radiation holds only if
essentially the whole virtual photon cloud around the electron is shaken off.
This can be expected to happen only if the amplitude of the electron's
transverse motion exceeds the minimum impact parameter $b_{\rm min} \approx
\gamma \lambda_C$ introduced in eq.~(\ref{eq63b}).  From eq.~(\ref{eq311a}) 
we see that the transverse amplitude obeys
\begin{equation}
R \approx \eta b_{\rm min}.
\label{eq312a}
\end{equation}
Thus, for $\eta$ less than one, the undulator radiation will be less than
full strength.  We readily expect that the intensity of weak radiation varies
as the square of the amplitude of the motion, so the estimate (\ref{eq304a})
should be revised as
\begin{equation}
{dU \over dt} \approx {\eta^2 e^2 c \gamma^2 \over \lambda_0^2},
\qquad (\eta \lsim 1).
\label{eq304b}
\end{equation}

The radiated power can be calculated exactly via the Larmor formula,
\begin{equation}
{dU \over dt} = {2 e^2 a^{\star 2} \over 3 c^3},
\label{eq312}
\end{equation}
where $a^\star = e E^\star/m$ is the acceleration of the electron in its 
instantaneous rest frame.  The electron is moving in a helix with its velocity
perpendicular to ${\bf B}_0$, so the electric field in the instantaneous rest
frame is $E^\star = \gamma\beta B_0 \approx \gamma B_0$.  Hence,
\begin{equation}
{dU \over dt} \approx {2 e^2 \gamma^2  \over 3 c} \left( {e B_0 \over m c} 
\right)^2
= {2 e^2 c \gamma^2 \eta^2  \over 3 \lambdabar_0^2},
\label{eq313}
\end{equation}
in agreement with the revised estimate (\ref{eq304b}).

In practice, $\eta \approx 1$ is the region of greatest  interest as it 
provides the maximum amount of radiation at the fundamental
frequency $\omega_C$.


\section{Transition Radiation}

As a charged particle crosses, for example,
a vacuum/metal boundary, its interaction with charges in the material results in
their acceleration and hence radiation, commonly called transition radiation.
The formation zone extends outwards from each boundary, with formation length 
given by eq.~(\ref{eq65}).  
The number of photons emitted as the particle crosses each boundary is given by
eq.~(\ref{eq63}) as $\alpha$ per unit frequency interval.  
If two boundaries are separated
by less than a formation length, interference effects arise that will not be 
considered here.
  
The minimum relevant transverse scale, $b_{\rm min}$, is the plasma 
wavelength $\lambdabar_p = c/\omega_p$, so the critical frequency is 
$\omega_C \approx \gamma \omega_p$, according to eq.~(\ref{eq63b}).
This is well into the x-ray regime.  While the characteristic angle
of transition radiation is $1/\gamma$, there is only a power-law falloff at
larger angles, and the optical transition radiation from an intense
beam of charged particles can be used to measure the spot size to accuracy of a
few $\lambda$ \cite{Chehab98a,Catravas}.

\section{\v Cerenkov Radiation}

When a charged particle moves with velocity $v$ in a dielectric medium,
\v Cerenkov radiation is emitted at those wavelengths for which $v > c/n$, 
where $n(\lambda)$ is the index of refraction. 
As the  particle approaches 
lightspeed in a medium, its electric field is compressed into a ``pancake",
which deforms into a cone when the the particle velocity exceeds lightspeed.
The particle outruns its electric field, which is freed as the \v Cerenkov
radiation.  The fields become identifiable as radiation after the particle
has moved a formation length, $L_0 = vt_0$, which is the distance over which
the electron pulls one wavelength ahead of the projection of the
wave motion onto the electron's direction.  The \v Cerenkov angle $\theta_C$ is
defined by the direction of the radiation, which is normal to the
conical surface that contains the electric field.  As usual,
$\cos\theta_C = c/nv = 1/n\beta$.  
The formation length is then
\begin{equation}
\lambda = vt_0 - {c \over n}t_0 \cos\theta_C 
= L_0 \sin^2 \theta_C. 
\label{eq80}
\end{equation}
Thus $L_0 = \lambda/\sin^2\theta_C$, and the photon spectrum per unit
path length from eq.~(\ref{eq63}) is
\begin{equation}
{dn_\omega \over dl} \approx {\alpha \over L_0} {d\omega \over \omega}
\approx {\alpha\sin^2\theta_C \over \lambda} {d\omega \over \omega}
\approx \alpha \sin^2\theta_C {d\omega \over c},
\label{eq81}
\end{equation}
as is well-known.  

The characteristic angle $\theta_C$ of \v Cerenkov radiation is essentially
independent of the Lorentz factor $\gamma$ of the charged particle, unlike
that for the other radiation processes considered here.  Correspondingly, the
characteristic transverse length $b$ associated with \v Cerenkov radiation
is also largely independent of $\gamma$. 
Rather, the region over which the \v Cerenkov radiation develops
has radius roughly that of the \v Cerenkov cone after one formation length,
\ie, $b = L_0 \cos\theta_C \sin\theta_C \approx
 \lambda / \tan\theta_C$.  This is large only near the \v Cerenkov threshold 
where the radiated intensity is very small.

That the formation radius for \v Cerenkov radiation is of order $\lambda$ is 
supported by an analysis \cite{Ginzburg} of a particle
moving in vacuum along the axis of a tube inside a dielectric medium; the
calculated \v Cerenkov radiation is negligible at wavelengths larger than the 
radius of the tube. 

 \v Cerenkov radiation is a form of energy loss for a particle
passing through a medium, and is related to so-called ionization loss
(see, for example, secs.~13.1-4 of \cite{Jackson}).  The latter is important
for frequencies higher than the ionization potential (divided by $\hbar$) of 
the medium, for which the index of refraction is typically less than one.
At frequencies that can cause ionization, the \v Cerenkov effect is
insignificant.  The transverse scale
of the ionizing fields grows with $\gamma$ due to relativistic flattening, but
shielding due to the induced dielectric polarization
of the medium results in an effective transverse scale 
$b \propto \sqrt{\gamma}$ for these fields when $\gamma \gg 1$.

\end{document}